\documentclass[sn-mathphys]{sn-jnl}% Math and Physical Sciences Reference Style
\usepackage{amsmath}
\usepackage{amssymb}
\jyear{2021}%

\theoremstyle{thmstyleone}%
\theoremstyle{thmstyletwo}%

\theoremstyle{thmstylethree}%

\newcommand{\revis}{}%{\color{red}}
\newcommand{\revisnew}{}%{\color{blue}}

\begin{document}

\title[Quantum equilibrium]{Relaxation to quantum equilibrium and the Born rule  in Nelson's stochastic dynamics}

\author[1]{\fnm{Vincent} \sur{Hardel}}\email{vincent.hardel@ipcms.unistra.fr}

\author[1]{\fnm{Paul-Antoine} \sur{Hervieux}}\email{paul-antoine.hervieux@ipcms.unistra.fr}

\author[1]{\fnm{Giovanni} \sur{Manfredi}}\email{giovanni.manfredi@ipcms.unistra.fr}

\affil[1]{{Universit\'e de Strasbourg, CNRS, Institut de Physique et de Chimie des Mat\'eriaux de Strasbourg}, \orgaddress{\city{67000 Strasbourg}, \country{France}}}

\abstract{Nelson's stochastic quantum mechanics provides an ideal arena to test how the Born rule is established from an initial probability distribution that is not identical to the square modulus of the wavefunction.
Here, we investigate numerically this problem for three relevant cases: a double-slit interference setup, a harmonic oscillator, and a quantum particle in a uniform gravitational field. For all cases, Nelson's stochastic trajectories are initially localized at a definite position, thereby violating the Born rule.
For the double slit and harmonic oscillator, typical quantum phenomena, such as interferences, always occur well after the establishment of the Born rule. In contrast, for the case of quantum particles free-falling in the gravity field of the Earth, an interference pattern is observed  \emph{before} the completion of the quantum relaxation. This finding may pave the way to experiments able to discriminate standard quantum mechanics, where the Born rule is always satisfied, from Nelson's theory, for which an early subquantum dynamics may be present before full quantum relaxation has occurred.
{\revisnew Although the mechanism through which a quantum particle might violate the Born rule remains unknown to date, we speculate that this may occur during fundamental processes, such as beta decay or particle-antiparticle pair production.
}
}



\maketitle

\section{Introduction}
Quantum mechanics (QM) has raised innumerable foundational questions since its formalization in the early twentieth century. Most of those questions arise from two ``weird" properties of QM, which single it out from earlier physical theories: (i) QM is an intrinsically probabilistic theory, meaning that its outcomes can only be predicted on average, and (ii) quantum probabilities do not follow the same rules as classical ones, inasmuch as in QM probability \emph{amplitudes} are additive, and not the probabilities themselves\footnote{It is possible to formulate QM in terms of ordinary probabilities, provided that these are allowed to take negative values (see, for instance, Ref. \cite{Manfredi2022} and references therein). This is another manifestation of the weirdness of quantum theory.}. This fact is encapsulated into Born's rule \cite{Born1926}, which defines quantum probabilities as the squared modulus of complex amplitudes.

The first of these properties was the source of much controversy at the dawn of QM, because earlier fundamental theories were all deterministic. Being capable of predicting with virtually perfect accuracy a physical event (e.g., an eclipse or the passage of a comet) was seen as the hallmark of a rigorous physical theory, the kind of achievement that gave Newton's and Maxwell's theories all their prestige. Besides, just a few years earlier, Boltzmann had shown how to bridge the gap between reversible macroscopic motion at the molecular level and irreversible heat and matter diffusion at the macroscopic scale. It was natural, then, to assume that also the randomness of QM could one day be explained in a similar fashion.

However, it is the second property that poses the hardest foundational questions -- and is also at the heart of the spooky action at a distance first highlighted in  the celebrated Einstein-Podolsky-Rosen (EPR) paper \cite{EPR1935}, and later confirmed in many experiments, mainly based on John Bell's extension  to spin states of the original EPR argument \cite{bell2004speakable}.
Born's rule is at the heart of these ``weird" features of QM and, for this reason, deserves some special attention. Indeed,  Born's rule stands alone in the mathematical machinery of QM, and is employed only when one needs to translate the abstract wavefunction into an actual prediction about probabilities of outcomes. We also note that, while the Schr\"odinger equation is linear in the wavefunction, Born's rule, which is quadratic, reinstates some nonlinearity into the theory.

It is well-known that in some nonlocal hidden-variable theories \cite{9formulations}, such as the Bohm-de Broglie version of QM (also known as Bohmian mechanics), the Born rule need not necessarily be satisfied\footnote{\revis Strictly speaking, actual ensembles in experiments only have a finite number of particles $N$, so that these theories always violate the Born rule. Here, we mean that the latter may be violated even in the limit $N \to \infty$.}. In the {\revis Bohm-de Broglie} mechanics \cite{Bohm1952}, if an ensemble of trajectories satisfies Born's rule at a certain initial time $t=0$, i.e. if $P(x,t=0) = \abs{\Psi(x,t=0)}^2$  (where $P$ is the probability density of the position variable $x$ and $\Psi$ is the wavefunction), then this property will always be satisfied for any subsequent time $t>0$. But the equations of the {\revis Bohm-de Broglie} mechanics remain perfectly valid also when one takes $P(x,t=0) \ne \abs{\Psi(x,t=0)}^2$, i.e., if Born's rule is violated. In that case, the two quantities $P(x,t)$ and $\abs{\Psi(x,t)}^2$ will remain distinct for all later times.

In the context of the {\revis Bohm-de Broglie} mechanics, Valentini \cite{valentini_signal-locality_1991} suggested that the Born rule is the analogue of thermal equilibrium in classical statistical mechanics. In the latter, non-equilibrium states are possible during transient evolutions, but the system eventually relaxes to its thermal equilibrium, given for instance by a Maxwellian probability  distribution. In the same fashion, Valentini postulated that the {\revis Bohm-de Broglie} distribution of positions may in general differ from that given by Born's rule, and only relaxes to it in a finite (albeit fast) timescale. Hence, the standard distribution that satisfies Born's rule corresponds to a sort of \emph{quantum equilibrium} defined by $P=\abs{\Psi}^2$, although quantum non-equilibrium states with $P \neq \abs{\Psi}^2$ may also exist during short transients (this is referred to as ``subquantum dynamics" by Valentini).
The possibility of finding signatures of subquantum dynamics in the primordial universe was also suggested \cite{Valentini2010,Underwood2015}.

{\revisnew Just like in standard statistical mechanics, quantum-equilibrium distributions are much more probable than non-equilibrium ones (they are \textit{typical}, in a technical sense\footnote{\revis For a definition of typicality in statistical mechanics,  see \cite{Lebowitz1993,Lebowitz1999}, and in the Bohm-de Broglie theory, see \cite{Durr1992,Durr2021}.}) and therefore should be observed most of the time, which is of course the case in all known experiments.
From a dynamical point of view, non-equilibrium distributions will typically converge to quantum equilibrium.
Earlier numerical simulations \cite{valentini2005dynamical} showed that relaxation to equilibrium is indeed observed, provided some coarse graining procedure is applied.}

An alternative, and perhaps more appropriate, avenue to study such convergence to quantum equilibrium is to resort to Nelson's stochastic quantization \cite{nelson_derivation_1966,bacciagaluppi1999nelsonian,Beyer2021}. As detailed in the next section, Nelson's dynamics is similar to the {\revis Bohm-de Broglie} mechanics, with the important difference that the equations of motion are not deterministic, but rather stochastic with a diffusion coefficient equal to $\hbar/2m$, where $\hbar$ is the reduced Planck constant and $m$ the mass. Nelson's theory reproduces standard QM when the Born rule is satisfied at the initial time. When this is not the case, the distribution $P$ will converge to the Born rule value $\abs{\Psi}^2$, without any need for an artificial  coarse graining procedure,  thanks to the stochastic nature of the dynamics.
Hence, Nelson's approach appears to be particularly adapted to investigate subquantum physics and the relaxation to quantum equilibrium.

{\revisnew Of course, one would also need to postulate a mechanism through which a quantum particle could find itself at quantum non-equilibrium. Although we do not have a theory for such a mechanism, we may conjecture that fundamental processes -- such as beta decay or particle-antiparticle pair production -- generate quantum particles that are, at least at the very early stages, out of quantum equilibrium.
Indeed, during such processes the quantum particles are created \textit{ex nihilo} and may not have had enough time to relax to the Born rule. We will not try to justify or explore any further this speculative conjecture. Our purpose here is merely to investigate what happens \textit{if}, for whatever reason, Born's rule is at some point violated.

Within this framework,} an important question is whether quantum thermalization occurs faster than any typical quantum effect, such as interference. If this is the case, it would mean that all typically quantum phenomena are ``equilibrium" phenomena and hence indistinguishable from standard QM. In the opposite case (i.e., quantum interference occurring before relaxation), one could hope to observe some anomaly in the interference  pattern due to subquantum corrections. If true, this would be an appealing prediction for future experiments.

In the present paper, we investigate this topic by means of numerical simulations of Nelson's stochastic dynamics, for three relevant cases: (i) a standard double-slit interference setup, (ii) a harmonic oscillator, and (iii) quantum particles in a gravity field, such as ultracold neutrons in the gravitational field of the Earth \cite{bouncing_neutron}.
The next section is devoted to a brief description of Nelson's approach to QM. In section \ref{Sec3}, we illustrate how to quantify the distance to quantum equilibrium and the relaxation towards it. Section \ref{Simulations}  includes the numerical results for the three physical systems mentioned above. Finally, conclusions are drawn in section \ref{Conclusion}.

\section{Nelson's stochastic quantization}\label{sec2}
In the {\revis Bohm-de Broglie} theory \cite{Bohm1952}, particles have a well-defined position $x(t)$, and their trajectories evolve according to a deterministic law of the type:
\begin{equation}\label{Bohmdebroglie}
            \frac{\dd x(t) }{\dd t}= u(x,t) ,
\end{equation}
where the velocity $u(x,t)$ is related to the phase of the wavefunction, which satisfies the standard time-dependent Schr\"odinger equation. In particular, writing the wavefunction in polar coordinates
$$\Psi(x,t) = R(x,t) \, \mathrm{e}^{iS(x,t)}, $$
where $R(x,t)$ is the amplitude and $S(x,t)$ is the phase, one has that {\revis $u=\hbar \partial_x S /m$}. Note that, in the present work, we will always consider one-dimensional problems.

In contrast, in Nelson's dynamics \citep{nelson_derivation_1966,bacciagaluppi1999nelsonian} the particles obey a Langevin equation
\begin{equation}\label{Nelson}
            \dd x(t) = b(x(t),t) \dd t + \dd W(t) ,
 \end{equation}
where $b(x(t),t)$ is the deterministic velocity and $W(t)$ is a stochastic Wiener process. The latter is characterized by a zero mean $\langle \dd W \rangle = 0$ and a finite variance
\begin{equation}
\langle \dd W^2 \rangle = D_Q \equiv  \frac{\hbar}{2m},
\label{Dq}
\end{equation}
with $D_Q$ the quantum diffusion coefficient. The origin of such  Brownian motion with diffusion coefficient $D_Q$ was not specified by Nelson, and here we just assume the presence of some universal force agitating all quantum particles.
We also note that similar stochastic theories have been discussed by Bohm and Hiley \cite{BohmHiley1989}, Peruzzi and Rimini \cite{peruzzi1996}, as well as Bohm and Vigier \cite{BohmVigier1954}.

In Nelson's theory, the total velocity $b(x,t)$ is written as the sum of two terms:
\begin{equation}\label{b_Nelson}
b(x,t) = \frac{\hbar}{m} \pdv{}{x}S(x,t) + 2D_Q \pdv{}{x}\ln R(x,t),
\end{equation}
where the first term (drift velocity) is proportional to the gradient of the phase and is identical to the velocity of the Bohm-de Broglie model, while the second term (osmotic velocity) depends on the amplitude $R$.

The wavefunction follows the standard Schr\"odinger equation $i\hbar \partial_t \Psi(x,t) = \hat H \Psi(x,t)$, with Hamiltonian $\hat H = \hat p^2/2m + \hat V(x,t)$. Hence, the phase $S$ obeys the following quantum Hamilton-Jacobi equation:
\begin{equation}\label{HamiltonJacobi}
\hbar \frac{\partial S}{\partial t} + {\hbar^2 \over {2m}} \left( \frac{\partial S}{\partial x} \right)^2 - {\hbar^2 \over {2m R}} \frac{\partial^2 R}{\partial x^2} + V = 0.
\end{equation}
Finally, the stochastic Langevin equation \eqref{Nelson} can also be  expressed as an equivalent Fokker-Planck equation for the probability density $P(x,t)$:
\begin{equation}\label{FokkerPlanck}
\frac{\partial P}{\partial t} + \frac{\partial }{\partial x} \left[b(x,t) P \right] = D_Q\, \frac{\partial^2 P}{\partial x^2}.
\end{equation}

In summary, Nelson's theory is encapsulated in the equations \eqref{Nelson} (stochastic process), \eqref{b_Nelson} (definition of the velocity), and \eqref{HamiltonJacobi} (quantum Hamilton-Jacobi).

When the initial particle distribution $P(x,0)$ is identical to the squared amplitude of the wavefunction $\abs{\Psi(x,0)}^2 = R^2(x,0)$,
Nelson's dynamics is equivalent to the standard quantum theory and reproduces the same results as the time-dependent Schr\"odinger equation. Like the Bohm-de Broglie theory, it can be seen as a nonlocal hidden variable theory, where the hidden variable is the position of the particles, but it differs from the {\revis Bohm-de Broglie} mechanics inasmuch as it is non-deterministic.
However, it is important to stress that, despite Eq. \eqref{Nelson} being a stochastic process, the whole Nelsonian dynamics is reversible in time \cite{nelson_derivation_1966}, as it should be to guarantee the equivalence with the Schr\"odinger equation. This can easily be seen from the Fokker-Planck equation \eqref{FokkerPlanck}, by noting that the osmotic velocity exactly cancels the diffusion term.

\section{Quantum equilibrium}\label{Sec3}

In the standard formulation of QM, the Born rule is a crucial postulate:  the probability density of finding a particle at a position $x$ at time $t$ is given by the squared modulus of the wavefunction $\abs{\Psi(x,t)}^2$. However, this postulate is not needed in the Nelson and Bohm-de Broglie formalisms, where the wavefunction is viewed as a field that guides the dynamics of the particles and is not necessarily linked to the probability of finding a particle in a certain region of space. Hence, it is perfectly consistent within these approaches to consider cases where $P(x,t) \neq  \abs{\Psi(x,t)}^2 $, in which case the predictions of standard QM would differ from those of the Nelson and Bohm-de Broglie theories.

As suggested by Valentini \cite{valentini_signal-locality_1991}, the Born rule may correspond to a situation of \emph{quantum equilibrium}, analogue to the thermal equilibrium of classical mechanics. According to this view, non-equilibrium states with $P(x,t) \neq  \abs{\Psi(x,t)}^2$ can exist, but  they relax to quantum equilibrium on a very short timescale,  so that they are difficult to observe in practice. Valentini developed these ideas in the context of the {\revis Bohm-de Broglie} mechanics which, being deterministic, requires some form of coarse graining to observe such relaxation \cite{valentini2005dynamical}. But in Nelson's theory the approach to equilibrium should occur more naturally, thanks to the stochastic nature of the motion. This fact was first analyzed in detail by Petroni and Guerra \cite{petroni_quantum_1995}, building on earlier work by Bohm and Vigier \cite{BohmVigier1954}, although the convergence to quantum equilibrium may not be proven in general for any initial condition and potential. More recently, Hatifi et al. \cite{hatif2018} have studied analytically and numerically the relaxation  to quantum equilibrium, in relation with the experiments of Couder et al. on bouncing oil droplets as an analogue of quantum motion \cite{Couder2005,Couder2006}.

The aim of the present work is to investigate, by means of numerical simulations, whether quantum thermalization occurs faster than any typical quantum effect, such as interference. In order to do so, one first needs to reconstruct the probability density $P(x,t)$ of he particles at each time. This is done by partitioning the space $x \in \mathbb{R}$ into bins of size $\Delta x$, such that each bin contains a sufficiently large number of particles, and constructing the corresponding histogram. The stochastic Nelson equation \eqref{Nelson} is solved using a  second-order Helfand-Greenside's method \citep{Greenside1981,Bayram2018,Rabitz1988}. In order to reduce the statistical noise, the simulations are repeated independently many times and the results are averaged to reconstruct the probability density.
In order to compute the velocity $b(x,t)$, we need to solve the Schr\"odinger equation to obtain the phase $S$ and amplitude $R$ of the wavefunction. In the three examples considered in this work, the solution could  be obtained analytically or semi-analytically, as detailed in the next section.

The probability density $P(x,t)$ must then be compared to the squared modulus of the wavefunction $\abs{\Psi(x,t)}^2=R^2$.
For this, we need to define a distance between these two quantities. Out of the many possibilities, one can use the
$L_p$ distance between two functions $f$ and $g$, defined as
\begin{equation}\label{Ln}
		    L_p\qty[f,g](t) = \sqrt[p]{\int_{-\infty}^{+\infty} \dd x~  \abs{f(x,t) - g(x,t)}^p}.
\end{equation}
In particular, the $L_1$ distance was advocated by Petroni and Guerra \cite{petroni_quantum_1995} as the appropriate tool to quantify the relaxation to quantum equilibrium.
The infinite distance $L_\infty$ can be seen as its limit when $p \to \infty$ and is given by
 \begin{equation}\label{Linf}
 L_\infty[f,g](t) = \max_{x}{\abs{f(x,t) - g(x,t)}}.
 \end{equation}
 Other criteria can also be defined, such as the entropy-like function used by Valentini \cite{valentini_signal-locality_1991}:
 \begin{equation}\label{Lh}
   H \equiv L_{H}\qty[f,g](t) = \int_{-\infty}^{+\infty}\dd x~ f(x,t) \ln\qty(\frac{f(x,t)}{g(x,t)}),
  \end{equation}
  which is related to the Kullback-Leibler divergence, also called relative entropy \cite{Joyce2011}.
 Taking $f=P$ and $g=\abs{\Psi(x,t)}^2$, all these distances vanish when the Born rule is satisfied, i.e. at quantum equilibrium.
 Of course, in order to estimate the relaxation time, it will be necessary to define a somewhat arbitrary threshold below which the distance is assumed to be practically zero.
Finally, using the entropy-like quantity \eqref{Lh}, Hatifi et al. \cite{hatif2018} were able to prove a H-theorem
which ensures that a generic probability distribution $P(x,t)$ converges to   $\abs{\Psi(x,t)}^2$ as $t \to \infty$ (with some caveats, as will be seen in the next section).

 \section{Simulation results}\label{Simulations}
The main question we try to answer in this work is whether quantum thermalization occurs faster than any other typical quantum effects, such as the appearance of interferences. If that were the case, it would mean that all  quantum phenomena are ``equilibrium" phenomena and hence indistinguishable from standard QM. In the opposite case, one could hope to observe some anomaly in the interference  pattern due to subquantum corrections, which  would be an appealing prediction for future experiments.

In this section, we will use the distance functionals defined in section \ref{Sec3} to estimate the time of relaxation to quantum equilibrium, and compare it with the time of appearance of quantum effects.
This problem will be investigated for three emblematic physical  systems: the double-slit experiment, the harmonic oscillator, and the evolution of a wavepacket in a linear potential representing the gravity field of the Earth.

\subsection{Double-slit experiment}

\begin{figure}[!ht]
		\begin{center}
			\includegraphics[scale=0.25]{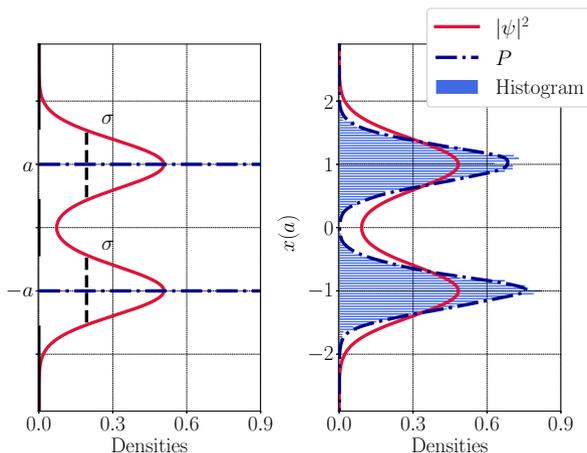}
		\end{center}
		\caption{Left panel: Initial densities for the wavefunction $\abs{\Psi(x,t)}^2$  (red continuous line) and the particles $P(x,0)$ (blue dashed line). Here, $P$ is the sum of two  Dirac delta functions centered at $\pm a$, while $\abs{\Psi(x,t)}^2$ is the sum of two Gaussians of width $\sigma = 0.3\, a$. Right panel: Same quantities at time $t=0.09\, \tau$, when quantum equilibrium is not yet attained.}
		\label{fig_initial_system}
\end{figure}

We consider a standard double-slit experiment, where the two slits have an aperture of width $\sigma$ and are separated by a distance $2a$, see figure \ref{fig_initial_system}. We shall use units in which $\hbar = m = a=1$, so that the only free parameter is the width $\sigma$ and actually represents the ratio $\sigma/a$. This choice also defines a timescale $\tau = ma^2/\hbar$ ($=1$, in these units).

In order to model the configuration of a double-slit experiment, we take an initial wavefunction that is the sum of two Gaussians of width $\sigma$ and centered at $x= \pm a$:
	\begin{equation}\label{psi_int}
 \Psi(x,0) = \frac{1}{\left[ 2\sqrt{\pi \sigma}\left( 1 + \mathrm{e}^{-a^2/\sigma^2} \right) \right]^{1/2}}\left( \mathrm{e}^{-(x+a)^2/2\sigma^2} + \mathrm{e}^{-(x-a)^2/2\sigma^2} \right) .
 	\end{equation}
As we want to investigate the relaxation to quantum equilibrium, the initial particle distribution should not satisfy the Born rule, i.e. $P(x,0) \neq \abs{\Psi(x,0)}^2$. Hence, we assume that all particles are concentrated at the same position, at the centre of each slit:
\begin{equation}\label{P_init_int}
	P(x,0) = \frac{\delta(x-a) + \delta(x+a)}{2},
\end{equation}
where $\delta$ denotes the Dirac delta function.
This initial configuration is plotted in figure  \ref{fig_initial_system} (left panel), while the right panel of the same figure shows both $\abs{\Psi(x,t)}^2$ and $P(x,t)$ at a later time when the system has evolved but has not yet reached the quantum equilibrium.
	
The free evolution of this initial wavefunction can be computed analytically \cite{Rabitz1988}, yielding the following square modulus at time $t$:
\begin{equation}\label{psi_t_int}
\begin{aligned}
\abs{\psi(x,t)}^2 = \frac{\sigma}{2\sqrt{\pi(\sigma^4+\frac{\hbar^2t^2}{m^2})\qty(1+\mathrm{e}^{-a^2/\sigma^2})}}\Bigg[\exp{-\frac{\sigma^2(x+a)^2}{\sigma^4+\frac{\hbar^2t^2}{m^2}}}\\
+\exp{-\frac{\sigma^2(x-a)^2}{\sigma^4+\frac{\hbar^2t^2}{m^2}}}+2\exp{-\frac{\sigma^2(x^2+a^2)}{\sigma^4+\frac{\hbar^2t^2}{m^2}}}\cos\qty(\frac{\frac{2\hbar tax}{m}}{\sigma^4+\frac{\hbar^2t^2}{m^2}})\Bigg].
\end{aligned}
\end{equation}
		
\begin{figure}[!ht]
			\begin{center}
				\includegraphics[scale=0.25]{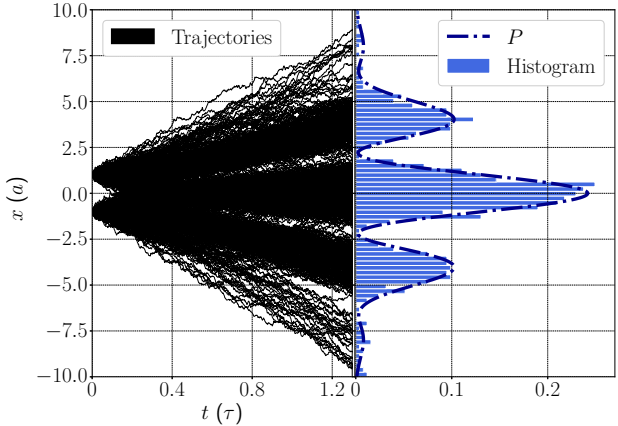}
			\end{center}
			\caption{Trajectories of $N=1000$ particles (left side, black curves)  initially distributed at the center of each slit. The histogram of the distribution of the positions (right side, blue segments) at the  end of the evolution is interpolated to obtain the corresponding density $P(x,t)$ (right side, dashed blue line).}
			\label{fig_trajectories}
\end{figure}

The particle density $P$ is obtained numerically by solving the stochastic Nelson equation (\ref{Nelson}) for a large number $N$ of trajectories.  In order to do so, one needs the expression of the velocity term $b$ that appears in the Nelson equation, which is obtained by injecting Eq. (\ref{psi_t_int}) into Eq. (\ref{b_Nelson}). We obtain \cite{Rabitz1988}:
\begin{equation}\label{b_t_int}
\begin{aligned}
b(x,t) = \qty(\Re+\Im)\Bigg( \frac{\hbar}{m} \frac{-\qty(\sigma^2-i\frac{\hbar t}{m})}{\sigma^4+\frac{\hbar^2t^2}{m^2}}\,\Bigg[(x+a)\exp{-\frac{\qty(\sigma^2-i\frac{\hbar t}{m})(x+a)^2}{2\qty(\sigma^4+\frac{\hbar^2t^2}{m^2})}}&\\
+(x-a)\exp{-\frac{\qty(\sigma^2-i\frac{\hbar t}{m})(x-a)^2}{2\qty(\sigma^4+\frac{\hbar^2t^2}{m^2})}}\Bigg]\\
\times\Bigg[\exp{-\frac{\qty(\sigma^2-i\frac{\hbar t}{m})(x+a)^2}{2\qty(\sigma^4+\frac{\hbar^2t^2}{m^2})}}+\exp{-\frac{\qty(\sigma^2-i\frac{\hbar t}{m})(x-a)^2}{2\qty(\sigma^4+\frac{\hbar^2t^2}{m^2})}}\Bigg]\Bigg) ,
\end{aligned}
\end{equation}
where $(\Re+\Im)$ denotes the sum of the real and imaginary parts of the expression between parenthesis.
Then, at each instant $t$, we construct a histogram of the particle positions, and finally interpolate the histogram to obtain the density $P(x,t)$. This procedure is illustrated in  figure \ref{fig_trajectories}.
	
	    \begin{figure}[!ht]
			\begin{center}
				\includegraphics[scale=0.25]{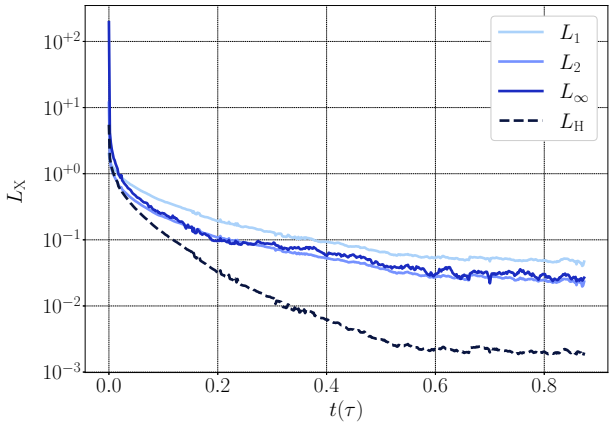}
			\end{center}
			\caption{Semi-logarithmic plots of the various functionals (see section \ref{Sec3}) used to quantify the distance between the probability density $P$ and the squared modulus of the wavefunction $\abs{\Psi}^2$, as a function of the time $t$ (in units of $\tau$), for $\sigma = 0.3 a$. }
			\label{distances}
		\end{figure}
		
Given the analytical expression of $\abs{\Psi}^2$ and the numerically-computed density $P$, it is possible to compare these two objects using the distances $L_\mathrm{X}$ defined in Section \ref{Sec3}. These quantities are represented as a function of  time in figure \ref{distances}, for the case $\sigma = 0.3 a$.		
For all cases, the distance between $P$ and $\abs{\Psi}^2$ decreases to zero for long times, signalling the convergence to the quantum equilibrium and the emergence of the Born rule. Due to numerical errors occurring during the computation of $P$, the minimal distance is never zero, but approximately $10^{-2}-10^{-3}$, depending on the adopted measure.  It is also interesting to note that the qualitative behavior is similar for all distances, so that they can be fitted with the same type of function in order to extract the relaxation time $\tau_\mathrm{q}$. Numerically, one can show that a good candidate for the fitting function is
		\begin{equation}\label{fit_ln_L}
		    L_\mathrm{X}(t) = \alpha_1 \exp\left(- \alpha_2 \mathrm{e}^{\alpha_3 t} \right),
		\end{equation}
 where $\alpha_1$, $\alpha_2$, and $\alpha_3$ are free fitting parameters, to be determined for each distance and each value of $\sigma$. From this expression, we define the quantum relaxation time $ \tau_\mathrm{q}$ as the time at which the tangent of the curve $L_\mathrm{X}(t)$ at $t=0$ intersects the abscissa axis, which gives: $\tau_\mathrm{q} = 1/(\alpha_2 \alpha_3)$.\footnote{Indeed, a Taylor expansion of Eq. \eqref{fit_ln_L} near $t=0$ yields: $ L_\mathrm{X}(t) \simeq L_\mathrm{X}(0)\,(1-\alpha_2 \alpha_3 t)$.}

\begin{figure}[!ht]
			\begin{center}
				\includegraphics[scale=0.25]{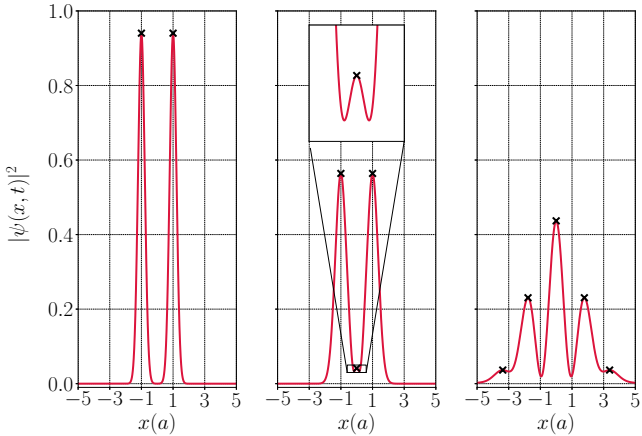}
			\end{center}
			\caption{Squared modulus of the wavefunction for $\sigma = 0.09a$, at times $t=0$ (left panel), $t=0.12\tau$ (middle panel), and $t = 0.6 \tau$ (right panel). Initially, only two peaks exist, one for each Gaussian wavepacket. At $t=0.12\tau$, a third peak has appeared between the two initial ones: this event defines the interference time $\tau_\mathrm{int}$. At later times, several new peaks appear and form the full interference pattern.}
			\label{fig_interference}
		\end{figure}

Next, we need a suitable definition of a ``typical" quantum time $\tau_\mathrm{int}$, defined as the time of appearance of quantum interferences, in order to compare it with the relaxation time $\tau_\mathrm{q}$. Interferences occur because the two initial Gaussian wavepackets spread in space, and after a certain time they overlap in the region between the two slits. As illustrated in figure \ref{fig_interference}, we define $\tau_\mathrm{int}$ as the time when the first maximum appears in between the two original wavepackets.
Further maxima appear at later times, until the full interference pattern is formed.
			
\begin{figure}[!ht]
			\begin{center}
				\includegraphics[scale=0.3]{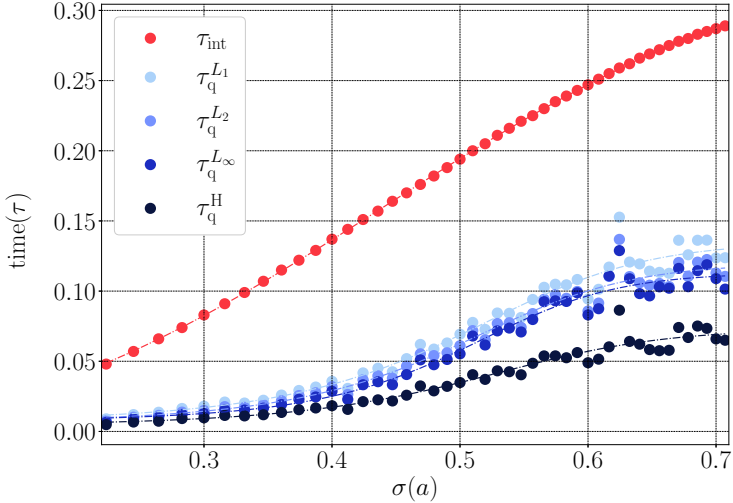}
			\end{center}
			\caption{Time of appearance of the interferences $\tau_\mathrm{int}$ (red dots) and times of convergence to quantum equilibrium $\tau_\mathrm{q}^{X}$ (shades of blue dots) associated with the different distances defined in section \ref{Sec3}, as a function of the initial width $\sigma/a$. All of the different times can be nicely fitted with a hyperbolic tangent function (dashed lines) of the type: $\tau_\mathrm{q}(\sigma) = \beta_1\tanh (\beta_2\sigma^2+\beta_3) +\beta_4$, where the $\beta_i$ are fitting parameters. For every value of $\sigma$ and for every distance $L_\mathrm{X}$, quantum equilibrium (Born's rule) is reached before the appearance of quantum interferences.}
			\label{fig_tau}
		\end{figure}

We now have all the elements to compare $\tau_\mathrm{q}$ and $\tau_\mathrm{int}$ for different values of $\sigma$.
The ratio $\sigma/a$ has to be smaller than unity to ensure that there is no significant overlap between the two Gaussian wavepackets at the initial time, but not too small because we want to ensure that $P$ and $\abs{\Psi}^2$ are  significantly different. Hence, we will consider values of $\sigma/a$ in the interval $[0.2, 0.7]$. The computed values of $\tau_\mathrm{int}$ and $\tau_\mathrm{q}$, for different distances $L_{X}$, are shown in figure \ref{fig_tau} as a function of the initial width $\sigma$.

The  important result of figure \ref{fig_tau} is that, whatever the value of $\sigma$, it is not possible to find a situation where the interference occurs before the system has converged to the quantum equilibrium. In other words, for the double slit experiment, all typically quantum physical phenomena  occur after the Born rule has been established. Or, to put it differently, the subquantum dynamics displays no quantum effects such as interferences.

A possible extension of the study presented in this section would be to consider three or more slits and check if it possibly increases the relaxation time beyond the quantum interference time. Experimental investigations in this direction have been performed recently \cite{many_slits,Cotter2017}. However, in the present work, we will rather focus on two other configurations: the harmonic oscillator and a linear potential truncated by a perfectly reflecting wall.

\subsection{Harmonic oscillator}\label{subsec:harmonic}

The harmonic oscillator is perhaps the most important and studied system in quantum mechanics and is crucial to the development of quantum field theory. It is both interesting in itself and a common approximation to many physical systems.
Here, we will further investigate the interplay between the establishment of the Born rule (quantum relaxation) and the appearance of typical quantum effects.

We consider the Schr\"odinger equation
 \begin{equation}
            i\hbar\pdv{}{t}\Psi(x,t) = \left(-\frac{\hbar^2}{2m}\pdv[2]{}{x} + \frac{1}{2}m \omega^2 x^2\right) \Psi(x,t) ,
 \end{equation}
where $m$ is  the mass of the particle and $\omega$ the frequency of the oscillator.
Normalizing space to $x_0 \equiv  \sqrt{\hbar/(m \omega)}$ and time to $t_0 \equiv 2/\omega$,
the Schr\"odinger equation becomes
\begin{equation}\label{schrod_harmonic}
            i\pdv{}{t}\Psi(x,t) = \left(-\pdv[2]{}{x} + x^2 \right) \Psi(x,t).
\end{equation}
This system of units amounts to taking $\omega = 2$, $\hbar=1$ and $m=1/2$, so that the quantum diffusion coefficient is  $D_Q = \hbar/2m = 1$ and the ground state energy $E_0 = m\omega^2/2 = 1$.

We want to study the convergence to the quantum equilibrium when  the initial particle probability density $P$ is given by a Dirac distribution centred at the bottom of the harmonic potential ($x=0$). The initial wavefunction is also a Gaussian of given width, but not necessarily the ground state of the system, hence it will display breathing oscillations while remaining Gaussian for all times.
A similar study, but only considering a ground state wavefunction for the Schr\"odinger equation, was performed by Hatifi et al. \cite{hatif2018}.

In practice, our initial condition is as follows:
 \begin{equation}
 \Psi(x,0) = \qty(\frac{B_0}{2\pi})^{\frac{1}{4}}
 \exp\left\{-\frac{B_0x^2}{4} + i\qty[\frac{A_0x^2}{2}+a_0]\right\} \quad\text{and}\quad P(x,0) = \delta(x),
  \end{equation}
where $A_0$, $B_0$ and $a_0$ are appropriate constants that define the wavefunction's width and phase.
At any time $t>0$, the wavefunction will keep the same functional form, so that it can be written as:
 \begin{equation}
 \Psi(x,t) = \qty(\frac{B(t)}{2\pi})^{\frac{1}{4}} \exp\left\{-\frac{B(t)x^2}{4} + i\qty[\frac{A(t)x^2}{2}+a(t)]\right\} ,
\end{equation}
with initial conditions $A(0) = A_0$, $a(0) = a_0$ and $B(0) = B_0$. Note that the ground state corresponds to
$A_0=a_0= 0$ and $B_0=2$.

Injecting this \textit{ansatz} into the Schr\"odinger equation (\ref{schrod_harmonic}), we obtain a system of first-order differential equations, where the dot denotes differentiation with respect to time:
        \begin{equation}\label{AaB}
            \left\{
                \begin{aligned}
                    \dot{A}(t) & = \frac{B(t)}{2} - 2A^2(t) -2, \\
                    \dot{a}(t) & = -\frac{B(t)}{2},\\
                    \dot{B}(t) & = -4A(t)B(t).
                \end{aligned}
            \right.
        \end{equation}
The solution to the above equations completely determines the wavefunction $\Psi(x,t)$, and hence the term $b(x,t)$ in Nelson's equation \eqref{b_Nelson}: $b(x,t) = [2A(t)-B(t)]x$, so that the Nelson equation can be written as
\begin{equation}
            \dd x(t) = [2A(t) - B(t)]\,x \dd t + \dd W(t).
\end{equation}
The corresponding Fokker-Planck equation can be obtained using the Kramers-Moyal expansion \cite{Kramers, Moyal} and reads as:
\begin{equation}\label{fokker_planck}
            \pdv{}{t}P(x,t) = \frac{\partial}{\partial x}\{-[2A(t) - B(t)]\,x\, P(x,t)\} + \pdv[2]{}{x}P(x,t).
\end{equation}
Supposing that the probability density is also Gaussian (which is an exact \textit{ansatz}):
        \begin{equation}
            P(x,t) = \sqrt{\frac{C(t)}{2\pi}} \,\exp\left(-C(t)\frac{x^2}{2}\right),
        \end{equation}
and injecting the above density into Eq. (\ref{fokker_planck}), one obtains that $C(t)$ should obey the following  equation
\begin{equation}\label{C}
            \dot{C}(t) = -2C(t) [2A(t) - B(t)]- 2C^2(t).
\end{equation}

The convergence to the quantum equilibrium can be studied by investigating the convergence of $C(t)$ to $B(t)$. To do so, we introduce the new variable $\gamma(t) = C(t)/B(t)$, which, from Eqs. (\ref{C}) and (\ref{AaB}), must be a solution of the Riccati equation
        \begin{equation}\label{gamma}
            \dot{\gamma}(t) = 2B(t)\gamma(t) [1-\gamma(t)].
        \end{equation}
Hence, one needs to first solve the system of equations (\ref{AaB}) to obtain $B(t)$ and then inject it into Eq.  (\ref{gamma}) in order to obtain $\gamma(t)$. The solution to Eq.  (\ref{gamma}) can be obtained pseudo-analytically and reads as \cite{handbook}:
        \begin{equation}\label{gamma_ana}
            \gamma(t) = 1 + \frac{\phi(t)}{2\int_0^t\dd \tau ~ B(\tau)\phi(\tau)} , \quad \text{with} \quad \phi(t) = \mathrm{e}^{-2\int_0^t\dd \tau B(\tau)}
        \end{equation}
with the initial condition $\gamma(0) = \infty$, which corresponds to the situation where $P$ is initially a Dirac delta function. Moreover, the system of equations \eqref{AaB} possesses the analytical solution  \cite{analytical_B}:
\begin{equation}\label{B(t)}
B(t) = \frac{8B_0}{B_0^2 + 4 - (B_0^4-4)\cos(4t)}.
\end{equation}

 \begin{figure}
            \centering
            \includegraphics[scale=0.3]{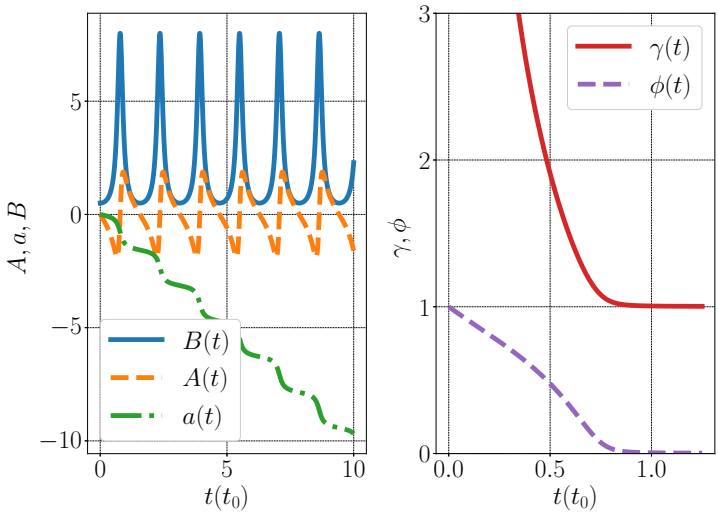}
            \caption{Left panel: Time evolution of the phase functions $A(t)$ and $a(t)$, and the inverse width $B(t)$ of the wavefunction $\Psi$.  $A(t)$ and $B(t)$ are periodic with period  $T = (\pi/2) t_0$, while $a(t)$ is monotonously decreasing, in accordance with the second equation \eqref{AaB}.
            Right panel: Time evolutions of the ratio $\gamma(t)=C(t)/B(t)$ and of the function $\phi(t)$ appearing in Eq. (\ref{gamma_ana});  $\gamma$ and $\phi$ converge respectively to unity and zero over a relaxation timescale denoted $\tau_\mathrm{q}$.}
            \label{fig_AaB}
  \end{figure}

In figure \ref{fig_AaB}, we present the solution of equations \eqref{AaB} and \eqref{gamma} for the initial conditions $A(0) =0, a(0) = 0, B(0) = 0.5$ and $\gamma(0) = \infty$, meaning, respectively, no initial phase, a wavefunction that is not the ground state of the harmonic oscillator, and a $\delta$-distributed probability $P(x,0)$.
The  phase function  $A$ and the width $B$ of the wavefunction are both periodic in time, with period $T = (\pi/2) t_0 = \pi/\omega$, equal to half the natural period of the harmonic oscillator $2\pi/\omega$ (this is because they are quadratic quantities in $x$).
In contrast, the ratio $\gamma=C/B$ relaxes to $\gamma=1$ over a timescale $\tau_\mathrm{q}$. When this has occurred, then both $P$ and $\abs{\Psi}^2$ are Gaussian functions of the same width and the Born rule is satisfied.

The purpose here is to compute $\tau_\mathrm{q}$ for different values of $B_0$, i.e. different initial widths of the wavefunction, and to check whether or not it is possible to find a situation where the period of quantum oscillations $T$ is shorter than the relaxation time $\tau_\mathrm{q}$.
In the following, we will consider different initial inverse widths $B_0$ of the wavefunction, from $B_0 = 0.125$ to $B_0 = 32$, corresponding to initial widths $\sigma_0 = \sqrt{2/B_0}$ from $0.25$ to $4$, in units of $x_0$. Note that, for the ground state, one has: $\sigma_0 = 1$ ($B_0 = 2$).

This can be done using several methods,  like arbitrarily defining a cutoff value, so that the relaxation time is defined as the time when $\gamma$ reaches such value. Here, we shall use a similar, but subtler, technique. We first compute the
root mean-square deviation of $\gamma$ over a sliding window in time \cite{moving_MSD}. We construct a window, centred at the data point $i$, which contains $n+1$ other data points between $i-n/2$ and $i+n/2$, and compute the mean square deviation $\Theta_i$ of $\gamma$ inside this window using the expression
 \[
 \Theta_i^2 =\frac{1}{n+1}\sum_{j=i-n/2}^{i+n/2}(\gamma_j -\bar{\gamma_i})^2,
 \]
where $\gamma_i = \gamma(t_i)$ and $\bar{\gamma_i} = \sum_{i-n/2}^{i+n/2} \gamma_j $ is the mean value of $\gamma$ inside the window. Typically, we take $n=10$. Hence, as $\gamma(t)$ approaches a constant value (here, $\gamma=1$), the function $ \Theta$ will tend to zero. By choosing a threshold $\theta$, one can define the relaxation time $\tau_\mathrm{q}$ as the time for which $\Theta < \theta$.

\begin{figure}[h!]
\centering
\includegraphics[scale=0.25]{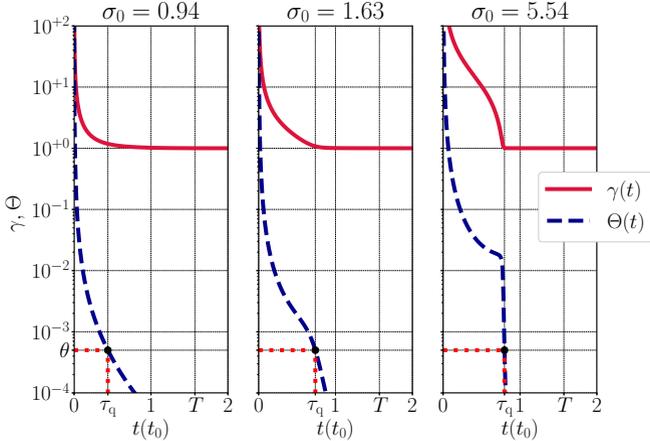}
\caption{Evolution of  $\gamma(t)$ (blue dashed curve) and its mean-square deviation $\Theta(t)$ (red solid curve) as a function of time (in units of $t_0$), for three different values of the initial wavefunction width $\sigma_0$. The cutoff value $\theta=5\times 10^{-4}$ is represented as a horizontal line which cuts the curve $\Theta(t)$ at $t=\tau_\mathrm{q}$, defining the relaxation time.
We note that $\tau_\mathrm{q}$ is  always smaller than the period  $T=(\pi/2)t_0$ of the harmonic oscillator (also represented on the abscissa axis), but increases when $\sigma_0$ increases.}
            \label{fig_gamma_Theta}
 \end{figure}

To visualize this procedure, the evolutions of $\gamma$ and $\Theta$ (dashed blue) are represented in figure \ref{fig_gamma_Theta}, for three  values of the initial width $\sigma_0 = 0.94$, $1.63$, and $5.54$. The convergence time is represented on the horizontal axis as the abscissa of the black dot, which is the point corresponding to $\Theta = \theta$, where in the present case $\theta = 5\times 10^{-4}$. For the different values of $\sigma_0$, the behavior of $\gamma(t)$ differs slightly, but the curve is always strictly decreasing, and no ambiguity arises for the determination of $\tau_\mathrm{q}$.

\begin{figure}[h!]
\centering
\includegraphics[scale=0.25]{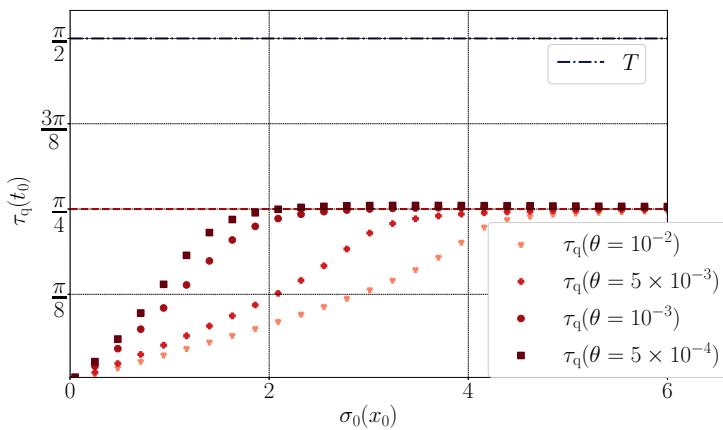}
\caption{Evolution of the quantum relaxation time $\tau_\mathrm{q}$ with respect to the initial width $\sigma_0$ of the wavefunction,  for different thresholds $\theta$, ranging from $1.0\times 10^{-2}$ to $5.0 \times 10^{-4}$ (shades of red dots).
For each threshold, the value of $\tau_\mathrm{q}$ increases with $\sigma_0$ and saturates at $\tau_\mathrm{q}=\pi/4$ (dotted red line). Hence, the convergence time is always at least twice as small as the quantum oscillator period $T = \pi/2$ (blue dashed line).}
            \label{fig_tau_rel_harmonic}
\end{figure}

 One may wonder about the dependence of the relaxation time on the threshold value $\theta$, but, as it appears in figure \ref{fig_gamma_Theta}, $\Theta$ decays fast close to the convergence time, so one can expect  this effect to be minor. To check this point, $\tau_\mathrm{q}$ was computed using different values of threshold, ranging from $\theta = 10^{-2}$ to $\theta = 5\times 10^{-4}$ and its dependence on the initial width $\sigma_0$ is plotted in figure \ref{fig_tau_rel_harmonic}.
 For every  threshold and for every value of $\sigma_0$, the relaxation time $\tau_\mathrm{q}$ is smaller than the  period of quantum oscillations $T$.
 In particular, we note the two limiting cases: (i) For $\sigma_0 \to 0$, then $\tau_\mathrm{q} \to 0$: this is rather natural, as it corresponds to the case where $P$ and $\abs{\Psi}^2$ already have the same vanishing width at $t=0$; (ii) For large $\sigma_0 $,  $\tau_\mathrm{q} \to \pi/4 = T/2$, in other words relaxation is completed in half an oscillation period.

 The limit  $\tau_\mathrm{q} \to \pi/4$, obtained for large initial dispersions, can be recovered analytically as follows. For small $B_0$, corresponding to large $\sigma_0 $, the function $B(t)$ becomes [see Eq.\eqref{B(t)}]:
 \[
B(t) \simeq\frac{2B_0}{1+\cos(4t)}=\frac{B_0}{\cos^2(2t)} ,
\]
so that, from Eq. \eqref{gamma_ana}: $\phi(t) \simeq \exp[-B_0\tan(2t)]$
which goes to zero when $t\to \pi/4$.

All in all, these results show that relaxation to quantum equilibrium (Born's rule) occurs much faster than an oscillation period of the quantum oscillator, and is completed at the latest over half such a period. As in the double-slit case, the system will always reach the quantum equilibrium before quantum phenomena become observable, preventing the possibility of observing a situation where the Born rule does not hold.

So far, we considered wavefunctions that are Gaussians, albeit not necessarily the ground state of the harmonic oscillator.
To end this section, we now turn to the case where $\Psi$ represents an excited state. In this case, the wavefunction possesses nodes (zeroes),  leading to singularities (asymptotes) in the velocity field $b(x,t)$, which becomes infinite at the location of the nodes. These singularities constitute infinite barriers that the trajectories cannot cross. For instance, for the first excited state of the oscillator, there is one singularity at $x=0$, where $\lim_{x \to 0^{\pm}} b(x) = \pm \infty$. Hence, a particle approaching zero from the right ($x >0$) will develop an ever increasing velocity directed in the positive $x$ direction, and will never manage to cross the origin. Similarly, for a particle approaching zero from the left ($x <0$).

\begin{figure}[h!]
\centering
\includegraphics[scale=0.25]{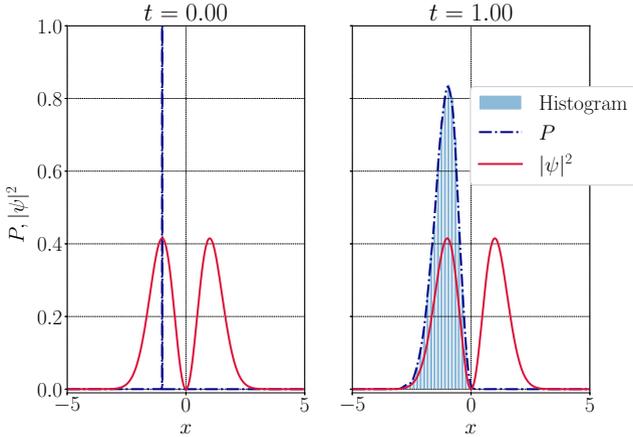}
\caption{Particle probability density $P(x,t)$ (dashed blue line) and squared wavefunction $\abs{\Psi}^2$ (red solid line) at times $t=0$ (left panel) and $t=1$ (right panel). Time is expressed in units of $t_0$ and space in units of $x_0$. The wavefunction corresponds to the first excited state of the harmonic oscillator. The initial particle distribution is a Dirac delta function centred at $x=-1$ and cannot cross the barrier located at the origin. The time step is $dt=10^{-4}$.}
\label{fig_forbidden}
\end{figure}

\begin{figure}[h!]
\centering
\includegraphics[scale=0.25]{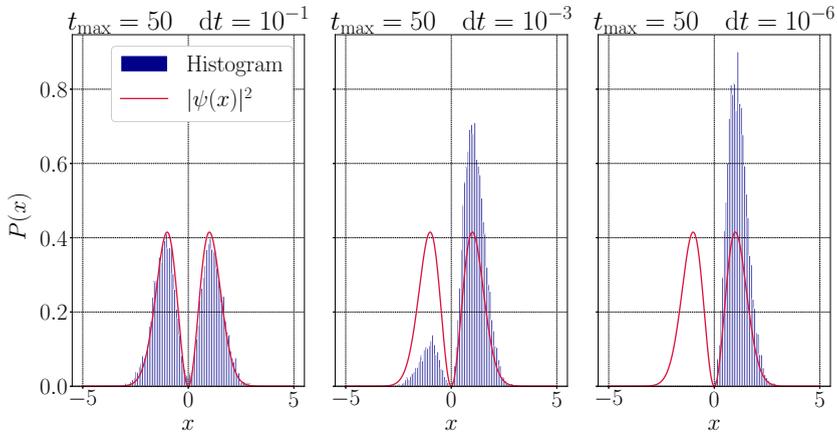}
\caption{Particle probability density $P(x,t)$ (blue histograms) and squared wavefunction $\abs{\Psi}^2$ (red solid line) at times $t=50$, for three values of the time step: $dt=0.1$ (left panel), $dt=10^{-3}$ (middle panel), and $dt=10^{-6}$ (right panel). Time is expressed in units of $t_0$ and space in units of $x_0$. The wavefunction corresponds to the first excited state of the harmonic oscillator and the particles are initially all located at $x=1$. For the smallest time step virtually no particles have crossed the barrier situated at $x=0$.}
\label{fig_influence-dt}
\end{figure}

This is illustrated in figure \ref{fig_forbidden}, where the initial distribution $P$ is a Dirac delta function located at $x=-1$, in the centre of the left lobe of the wavefunction density. At $t=1$ (right panel), the initial particle distribution has considerably spread, but it has not crossed the barrier at $x=0$.
We note that this result is in disagreement with a similar simulation of Hatifi et al. \cite{hatif2018}, who found numerically that the barrier is eventually crossed and full relaxation is observed. Nevertheless, some important differences exist: firstly, Hatifi et al. \cite{hatif2018} simulate a single trajectory and appeal to the ergodic theorem to reconstruct the particle density $P$; secondly, their final simulation time $t_{\rm final} = 1000$  is much longer than ours (this is because they have to average on time slices to compensate for the presence of a single trajectory).
But the main difference is in the time step, which is $dt=0.01$ in their simulation and $dt=10^{-4}$ in ours. Indeed, if the time step is large enough, the particle can sometimes cross the barrier, because it cannot ``see" it during times shorter than $dt$.
This is confirmed by three long-time simulations using different values of $dt$ (see figure \ref{fig_influence-dt}), which show that, as the time step decreases, fewer and fewer particles cross the barrier. Hence, in the limit $dt \to 0$, no crossings should be observed.

\begin{figure}[h!]
\centering
\includegraphics[scale=0.25]{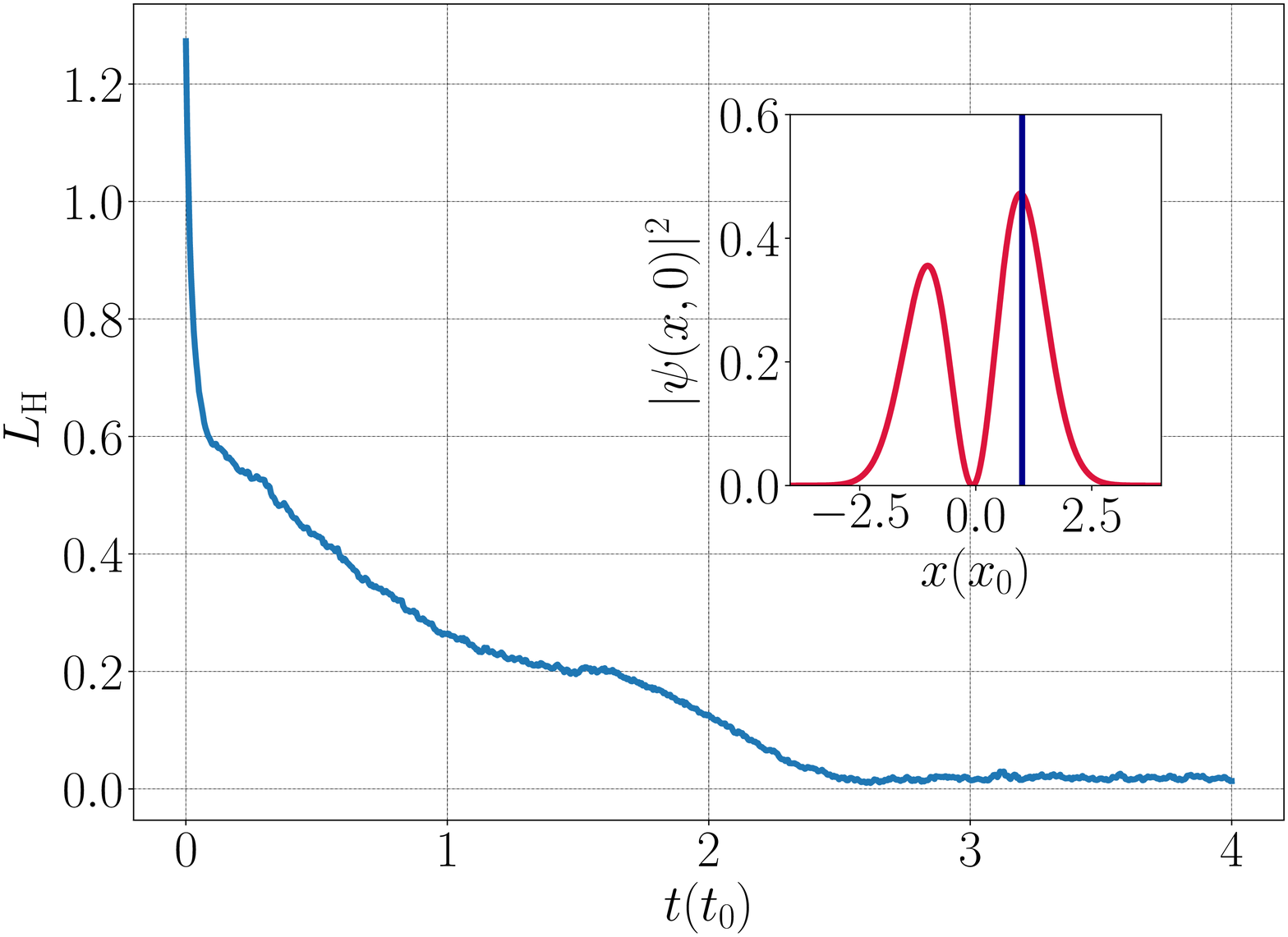}
\caption{Time evolution of the distance $L_\mathrm{H}(t)$ for an initial state that is a superposition of the ground state $\Psi_0(x)$ and the first excited state $\Psi_1(x)$: {\revisnew $\Psi(x,0) = \sin(0.1^{\circ}) \Psi_0(x) + \cos(0.1^{\circ}) \Psi_1(x) $} (the corresponding density is shown in the inset). Initially, the particles are localized at $x=1$ (blue vertical line in the inset). Time is expressed in units of $t_0$ and space in units of $x_0$. Relaxation is completed for $t=\tau_\mathrm{q} \approx 2.8 t_0$, shorter than the oscillator period $2\pi/\omega = \pi t_0$. }
\label{fig_superposition}
\end{figure}

The result of figure \ref{fig_forbidden} may seem in contradiction with what was claimed earlier, namely that the  relaxation time $\tau_\mathrm{q}$ is smaller than any typical quantum timescale. In figure \ref{fig_forbidden}, relaxation never occurs, so effectively $\tau_\mathrm{q} \to \infty$.
To better understand this issue, we have performed one further simulation (see figure \ref{fig_superposition}) for an initial wavefunction that is equal to the first excited state $\Psi_1(x)$, plus a small perturbation proportional to the ground state $\Psi_0(x)$: {\revisnew $\Psi(x,0) =  \cos(0.1^{\circ}) \Psi_1(x) +\sin(0.1^{\circ}) \Psi_0(x)$ (note that $\sin(0.1^{\circ}) \approx 0.0017 \ll 1$)}. In this case, relaxation takes place again and occurs on a timescale  $\tau_\mathrm{q} \approx 2.8 t_0$, shorter than the oscillator period $2\pi/\omega = \pi t_0$ (remember that $\omega=2/t_0$ in our units).
In summary, the relaxation time $\tau_\mathrm{q}$ is indeed always smaller than the typical oscillator timescale, except in the special case of an initial wavefunction that is an eigenstate of the system and possesses one or more nodes.

\subsection{Uniform gravity field}\label{subsec:gravity}
\subsubsection{Ultracold neutron experiments}
Let us now consider the case of a particle in a constant field, like the one generated by the gravitational attraction of the Earth. This type of problems are motivated by ongoing experiments on the gravitational response of antimatter, in which anti-hydrogen atoms fall in the gravity field of the Earth and are annihilated at the lower surface of the device \cite{ALPHA2013,GBAR2012}. By measuring the duration of the fall, it will  be possible to estimate the gravitational acceleration of antimatter $\bar g$, and check whether it is identical to that of standard matter $g$.

When the quantum nature of the anti-hydrogen atoms is taken into account, more subtle phenomena can arise, leading to the quantum reflection of the atoms at the surface through the Casimir-Polder potential \cite{Polder_Potential} and the subsequent formation of an interference pattern. Exploiting this effect can considerably improve the estimation of $\bar g$, because of the great precision with which frequency differences can be measured \cite{these1, these2, these3}.

Similar experiments were  performed over two decades ago using free-falling ultracold neutrons confined between a lower reflecting mirror and an upper absorbing surface \cite{bouncing_neutron}, and  led to the observation of the quantized gravitational energy levels of the neutrons. These techniques were further used to realize high-precision gravity-resonance spectroscopy studies on ultracold neutrons \cite{Jenke_NP11}, which were recently exploited to search for anomalous gravitational interactions \cite{Jenke_PRL14}. Gravitational experiments that use cold hydrogen atoms are also envisaged \cite{killian2023grasian}

Here, we will focus on the relaxation to quantum equilibrium of a quantum particle (typically, a neutron) falling in the gravitational field of the Earth from a height $h$. The initial wavefunction is a Gaussian of width $\zeta$ centered at $x=h$, where $x$ is the coordinate representing the altitude with respect to the lower reflecting mirror, whereas the particles are initialized as a Dirac delta function at the same height $h$. After bouncing on the mirror, the wavefunction develops quantum interferences. Our purpose will be again to investigate whether quantum relaxation and the establishment of the Born rule occurs before or after the formation of the quantum interference pattern.

\subsubsection{Gravitational quantum states}
Assuming a constant gravitational force at the surface of the Earth, the corresponding gravitational potential is $m g x$, where $m$ is the mass of the neutron, $g$ the free-fall acceleration, and $x$ the altitude with respect to the reflecting mirror, located at $x=0$. The corresponding wavefunction is a solution of the time-dependent Schr\"odinger equation
    \begin{equation}\label{schrodinger_gravity}
        i\hbar\pdv{}{t}\Psi(x,t) = \left(-\frac{\hbar^2}{2m}\pdv[2]{}{x}+ mgx \right) \Psi(x,t),
    \end{equation}
with boundary conditions $\Psi(x=0,t) = \Psi(x \to \infty,t) = 0$, for all times. The system is then bound and admits a discrete set of eigenstates. The initial wavefunction is given by
    \begin{equation}\label{initial_wavefunction_gravity}
        \Psi(x,0) = \Theta(x)\frac{1}{(2\pi\zeta^2)^{\frac{1}{4}}} \, \exp\left[-\frac{(x-h)^2}{4\zeta^2}\right] ,
    \end{equation}
with $\Theta(x)$ the Heaviside function, ensuring that the wavefunction is strictly zero for $x \le 0$. We choose $\zeta \ll h$, so that the wavefunction is correctly normalized.

\begin{figure}[h!]
        \centering
        \includegraphics[scale = 0.26]{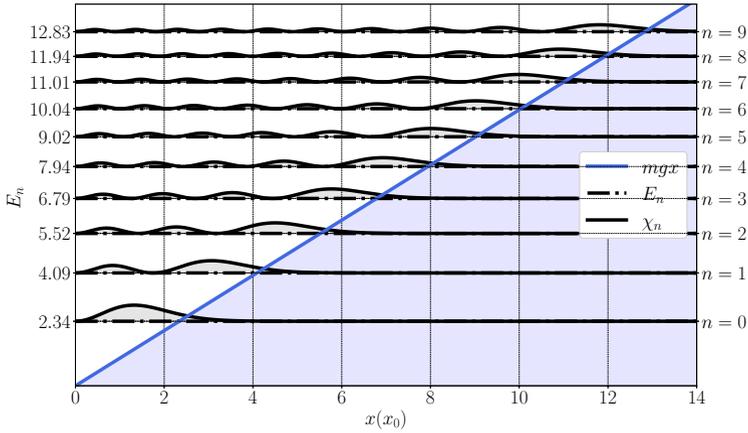}
        \caption{Representation of the first ten gravitational quantum states $\chi_n$ (red solid lines), which are given by the same Airy function $\mathrm{Ai}(x)$ shifted of an amount equal to $E_n$, where $E_n$ is the $n$-th energy eigenvalue; see Eq. \eqref{eigenfunctions_gravity} for the full formula. The horizontal axis represents the altitude $x$, in units of $x_\mathrm{0}$. The blue line represents the gravitational potential $mgx$. }
        \label{fig_airy}
    \end{figure}

The eigenstates $\chi_n$ of the problem are obtained by solving the stationary Schr\"odinger equation
\begin{equation}\label{schrod_independent_gravity}
        \qty(-\frac{\hbar^2}{2m}\pdv[2]{}{x}+mgx)\chi_n(x) = E_n\chi_n(x).
 \end{equation}
 We further define dimensionless units of length, energy and time as follows:
 \begin{equation}
  x_\mathrm{0} = \qty(\frac{\hbar^2}{2m^2g})^{\frac{1}{3}}, \quad  \epsilon_\mathrm{0} = m g x_\mathrm{0} = \qty(\frac{\hbar^2mg^2}{2})^{\frac{1}{3}}, \quad t_\mathrm{0} = \frac{\hbar}{\epsilon_\mathrm{0}} = \qty(\frac{2\hbar}{mg^2})^{\frac{1}{3}} .
  \label{eq:units}
 \end{equation}
Using these units, the eigenfunctions read as:
     \begin{equation}\label{eigenfunctions_gravity}
        \chi_n(x) = \Theta(x)\,\frac{\mathrm{Ai}\qty(x - E_n)}{\mathrm{Ai}'(-E_n)},
    \end{equation}
where $\mathrm{Ai}(x)$ denotes the first Airy function and $\mathrm{Ai}'(x)$ its derivative.
Because the eigenenergies are obtained by imposing $\chi_n(0) = 0$, they correspond to the zeros of the Airy function $\mathrm{Ai}$, which are well-known and have been tabulated \cite{Suda2022}. It is also possible to convert each $E_n$ to a corresponding ``eigenaltitude" $h_n$ above the mirror surface, by setting $E_n$ equal to the potential energy $mg h_n$, leading to: $h_n = E_n/mg$.
The presence of an upper absorbing plate ensures that only a finite number $n_\mathrm{max}$ of eigenstates can be present simultaneously in the device.
The first ten eigenfunctions are represented in figure \ref{fig_airy}, together with the eigenenergies/eigenaltitudes and the gravitational potential $mgx$.

Using the eigenbasis \eqref{eigenfunctions_gravity}, the solution to the Schr\"odinger equation \eqref{schrodinger_gravity} can be written as
    \begin{equation}\label{time_wavefunction_gravity}
        \Psi(x,t) = \sum_{n=0}^{n_\mathrm{max}} c_n \chi_n(x)\mathrm{e}^{-i E_n t} ,
    \end{equation}
 where the $c_n$ are the coefficients of the expansion \cite{Valle}. Their expression can be obtained semi-analytically under the assumption that the width $\zeta$ of the wavepacket is small compared to its altitude $h$ \cite{Crepin}:
    \begin{equation}\label{c_n_gravity}
        c_n = \frac{(8\pi\zeta^2)^{\frac{1}{4}}}{\mathrm{Ai}'(-E_n)}\mathrm{Ai}\qty(h-E_n+\zeta^4)\exp{\zeta^2\qty(h-E_n+\frac{2}{3}\zeta^4)}.
    \end{equation}
Some details of the derivation are given in the Appendix \ref{apdxAA}.

\subsubsection{Relaxation to quantum equilibrium}
\begin{figure}[h!]
        \centering
        \includegraphics[scale=0.26]{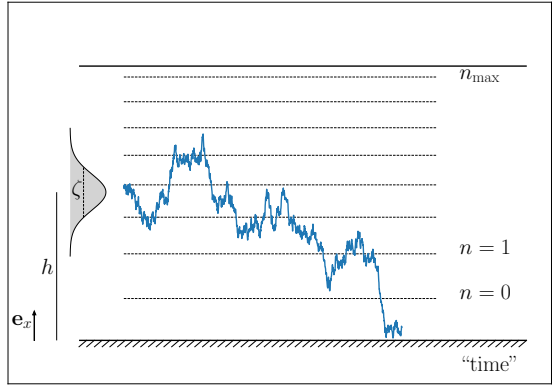}
        \caption{Schematic view of the physical system under study. The initial wavefunction (grey curve on the left) is a Gaussian of width $\zeta$, centered at an altitude $h$ from the mirror (hatched horizontal line at the bottom). The different eigenaltitudes (dashed horizontal lines) are represented for $n = 0,1,\cdots n_\mathrm{max}$, where $n_\mathrm{max}$ is the highest-energy state allowed by the upper  absorbing plate. The trajectory of a typical particle (blue line), initially located at $x=h$, shows the presence of bounces, not only at the level of the mirror, but also in correspondence of the various eigenaltitudes.}
        \label{fig_initial_system_gravity}
    \end{figure}

In order to investigate the relaxation to quantum equilibrium, we take an initial probability distribution $P$ that does not follow the Born rule, but is rather given by a Dirac delta function: $P(x,0) = \delta(x-h)$, so that all  particles are at the same altitude $h$ from the mirror. In the forthcoming simulations the altitude varies from $h=1.50$ -- which is lower than the ground-state eigenaltitude ($h_0=2.34$) -- to $h=5$.
The width of the initial  wavefunction is fixed and equal to $\zeta = 0.09$.
A schematic representation of the initial system, along with a typical random trajectory obtained by solving Nelson's stochastic equation, is shown in figure \ref{fig_initial_system_gravity}.

\begin{figure}[h!]
            \centering
            \includegraphics[scale=0.28]{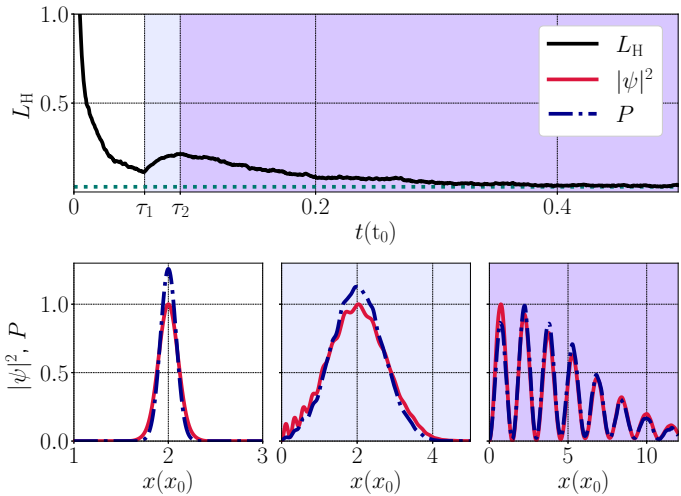}
            \caption{Upper panel: Time evolution of the distance $L_\mathrm{H}(t)$ for an initial state with $h=1.5$ and $\zeta=0.09$. The shaded colours represent the three different phases of the evolution described in the main text. The two vertical dashed lines show the times $\tau_1$ and $\tau_2$ between which the $L_\mathrm{H}$ distance increases. The dashed  horizontal line corresponds to the level below which $L_\mathrm{H}$ cannot go, for reasons due to the numerical integration {\revis (errors due to the finite number of particles and the interpolation method)}. Full convergence -- hence establishment of the Born rule -- is achieved for a relaxation time $\tau_\mathrm{q} \approx 0.5$, significantly larger than $\tau_2$.
            Lower panels: Squared modulus of the wavefunction $\abs{\Psi}^2$ (red solid curve) and particle distribution $P$ (blue dashed curve) at three different times, $t=0.005$ (left), $t=0.07$ (middle), and $t=0.5$ (right) (in units of $t_0$), corresponding to the three regions visible in the upper panel. }
            \label{fig_system_neq}
        \end{figure}

The $L_\mathrm{H}$ distance as a function of time is shown in figure \ref{fig_system_neq} (upper panel) and displays a peculiar behaviour. First, it decreases rather abruptly until a time $\tau_1$, then it increases up to time $\tau_2$, and finally decreases again for $t> \tau_2$.
In order to understand this behaviour, the squared modulus of the wavefunction $\abs{\Psi}^2$ and the probability density $P$ are also shown in figure \ref{fig_system_neq} (lower panels) for three different times $t=0.005$,  $t=0.07$ and $t=0.5$, corresponding to  three different  phases of the evolution: (i) $t < \tau_1$, (ii) $\tau_1 < t < \tau_2$, and (iii) $t > \tau_2$.
During the first phase, both $\abs{\Psi}^2$ and $P$ remain approximately Gaussian and their distance is progressively reduced, as it was found for the harmonic oscillator in section \ref{subsec:harmonic}. However, after $\tau_1$, interferences start building up in $\abs{\Psi}^2$, but not in $P$, so that the distance between such two functions increases again. For $t > \tau_2$, the interference pattern is fully formed and the particle distribution again converges towards $\abs{\Psi}^2$.

Finally, for even longer times, of the order of the relaxation time $\tau_\mathrm{q} \approx 0.5$, the $L_\mathrm{H}$ distance goes to zero and the Born rule is eventually satisfied (figure \ref{fig_system_neq}, upper panel).
Hence, it appears that some quantum interference phenomena do occur before the quantum relaxation is fully completed, in particular during the intermediate phase where $\tau_1 < t < \tau_2$, where the distributions $\abs{\Psi}^2$ and $P$ start diverging again.
During that phase, the interference pattern forms too quickly for the particle distribution to catch up with the wavefunction.
This type of effect was not observed in the two other situations (double slit and harmonic oscillator) that were analysed earlier in the present work.

\begin{figure}[h!]
\centering
\includegraphics[scale=0.25]{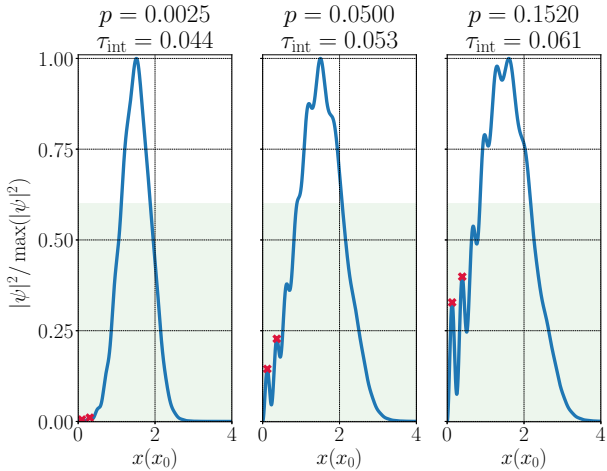}
\caption{Normalized squared modulus of the wavefunction as a function of the distance $x$ from the lower mirror, for an initial height $h=1.50$ (in units of $x_0$) and three values of the prominence: $p=0.0025$ (left panel), $p=0.05$ (middle panel), and  $p=0.152$ (right panel). Interference is said to occur when at least two peaks are present in the green shaded region and have a prominence higher than $p$.  The peaks are highlighted by a red cross on the curves. The corresponding interference time $\tau_\mathrm{int}$ depends on the chosen value of $p$ and is also indicated on the figure.}
 \label{fig_peaks_prominence}
\end{figure}

In order to show that the time $\tau_1$ (when the distance between $\abs{\Psi}^2$ and $P$ starts increasing again) actually coincides with the time of appearance of the early interference pattern $\tau_{\rm int}$, we need a recipe to estimate the latter. The procedure runs as follows.
First, we normalize the squared modulus of the wavefunction so that its maximum is equal to unity, and search for extrema in the region  $0<\abs{\Psi}^2/\max\abs{\Psi}^2<0.6$, thus focussing on the tail of the wavefunction (shaded green area in figure \ref{fig_peaks_prominence}).
Then, we define the prominence of a peak as the height between two neighbouring extrema (a maximum and a minimum).
We consider that interference occurs when at least two peaks have appeared with prominence larger than a threshold value $p$.
This defines the appearance time of the interference pattern, $\tau_\mathrm{int}$. This procedure is illustrated in figure \ref{fig_peaks_prominence}, where the wavefunction at the interference time is plotted for three values of $p$.

\begin{figure}[h!]
            \centering
            \includegraphics[scale = 0.25]{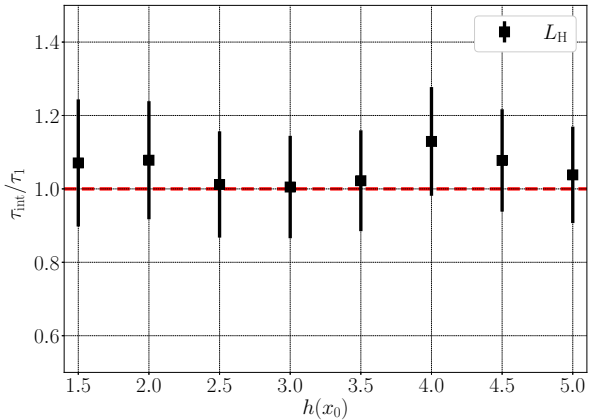}
            \caption{Ratio of the interference time $\tau_{\rm int}$ and the time of increase of the $L_\mathrm{H}$ distance $\tau_1$ (black squares) for different altitudes $h$ and an intermediate value of the prominence, $p=0.05$, see figure \ref{fig_peaks_prominence}. The ``error bars" are obtained using the upper and lower values $p=0.0152$ and $p=0.0025$. All ratios are close to unity, indicating that the two times relate to the same physical phenomenon.}
            \label{fig_ratio}
\end{figure}

Now, we can compare the interference time  $\tau_\mathrm{int}$ with the time $\tau_1$ at which the $L_\mathrm{H}$ distance starts increasing. The result is plotted in figure \ref{fig_ratio}, including error bars accounting for different choices of the prominence $p$. As expected, these two times are very similar, confirming that the increasing distance between $\abs{\Psi}^2$ and $P$ between $\tau_1$ and $\tau_2$ is due to the formation of an early interference pattern in the former, but not in the latter.

In summary,  simulations of a quantum particle falling in a uniform  gravitational field have shown that quantum interference phenomena could indeed be observed before the Born rule is satisfied, in contrast to what was found for the double slit and harmonic cases.
This opens the way to possible experimental verifications of the Born rule using gravitational quantum states of ultracold neutrons \cite{bouncing_neutron} or hydrogen atoms \cite{killian2023grasian}, which, in the case of neutrons, have reached extremely high accuracy levels \cite{Jenke_PRL14}.
We recall that we expressed our results in units of $x_0 = 5.87 \rm \mu m$ for distances and $t_0 = 1.09 \,\rm ms$ for times, see Eq. \eqref{eq:units}. Hence, for the case of figure \ref{fig_system_neq}, a significant discrepancy from the Born rule should still be observable around $t \approx 0.2t_0 \approx 0.2 \rm ms$, if all neutrons were initially perfectly localized at an altitude $h = 1.5 x_0 \approx 8.8 \rm \mu m$. This level of accuracy in the time resolution should be attainable with current  experimental setups.

\section{Conclusion}\label{Conclusion}

The Born rule was introduced by Born in 1926 in order to provide an interpretation of the wavefunction that appears in the Schr\"odinger equation. Interestingly, in the original paper by Born \cite{Born1926}, the rule appears in a note added in proofs, and is expressed in words rather than mathematically.\footnote{The footnote reads as \cite{Born1926}: \textit{Anmerkung  bei der Korrektur: Genauere Uberlegung zeigt, da{\ss} die Wahrscheinlichkeit  dem  Quadrat der  $\Psi$ proportional  ist}. (Note added in proofs: More careful consideration shows that the probability is proportional to the square of $\Psi$).}
Such simple rule stands alone with respect to the mathematical machinery of quantum mechanics, but is of course extremely important, as it bridges the gap between the abstract mathematical theory and the interpretation of actual experiments.

A question that has been raised by several researchers is whether the Born rule should be considered as fundamental, or rather an approximation. In particular, Valentini \cite{valentini_signal-locality_1991,valentini2005dynamical}  suggested that the Born rule plays the same role as thermal equilibrium in classical statistical mechanics. Just like an out-of-equilibrium  classical system quickly relaxes towards a Maxwell-Boltzmann equilibrium, a quantum system may exist in a ``subquantum" state where the Born rule is not satisfied. We always observe the validity of the Born rule only because this relaxation to quantum equilibrium is extremely fast.

Nelson's stochastic version of quantum mechanics provides an ideal arena to test such subquantum dynamics, as it allows to initialize the system in an out-of-equilibrium state that does not respect the Born rule. Due to the random nature of Nelson's dynamics, the Born rule is quickly attained over a timescale that depends on the system under study. (The same is true for the Bohm-de Broglie theory, but the latter being deterministic, it requires some sort of coarse graining in order to recover Born's rule).

In the present work, we have investigated numerically this relaxation to quantum equilibrium for three relevant cases: a standard double-slit interference setup, a harmonic oscillator, and a quantum particle in a uniform gravity field, such as ultracold neutrons in the gravitational field of the Earth. For all cases, the Nelson stochastic trajectories are initially localized at a definite position, thereby violating the Born rule.

For the double slit and harmonic oscillator, we found that typical quantum phenomena, such as interferences, always occur well after the establishment of the Born rule. In contrast, for the case of quantum particles free-falling in the gravity field of the Earth, an interference pattern is observed  \emph{before} the completion of the quantum relaxation.
The different behavior in the latter case is likely to arise from the nonlinearity induced by the reflecting mirror. If that is the case, a similar behaviour should be observed for generic non-quadratic Hamiltonians.

These findings may pave the way to experiments that are capable of discriminating standard quantum mechanics,
 where the Born rule is always verified, from Nelson's theory, for which an early subquantum dynamics may be present before full quantum relaxation has occurred.
{\revis
One may argue that particles in our labs had  a long and violent astrophysical history since the Big Bang, with ample time to relax to quantum equilibrium, so that it would be extremely difficult to observe any deviations from the Born rule at the present epoch. This is the line of argument followed by Valentini \cite{Valentini1996} in the context of the Bohm-de Broglie theory.

However, {\revisnew one might speculate} on different scenarios. For instance, we could  think of a decay-type experiment (beta or alpha decay, neutron or proton emission, etc.) in which a quantum particle (electron, positron, helium nucleus, neutron, proton...) is created from a fundamental process arising -- for instance, but not exclusively -- from the weak interaction. In this case,
 {\revisnew the particle might be born} in a non-equilibrium situation where Born's rule has not had enough time to be established. Another example is the creation of a particle-antiparticle pair (e.g., electron-positron) from a photon. This occurs in nuclear physics when a high-energy photon interacts with the nucleus, enabling the production of an electron-positron pair without violating the conservation of momentum.  Just after the pair creation, the electron or positron should be in a non-equilibrium state.
 {\revisnew Of course, these are somewhat speculative proposals}, but the findings put forward in this work at least suggest a viable way to test the existence of a subquantum dynamics in laboratory experiments.
}

\begin{appendices}

\section{Derivation of the $c_n$ coefficients}\label{apdxAA}

We sketch here the procedure used in Ref. \cite{these3} to decompose the wavefunction
 \[
        \Psi(x,0) = \frac{1}{(2\pi\zeta^2)^{1/4}} \, \exp\left[-\frac{(x-h)^2}{4\zeta^2} \right]
 \]
 on the basis of the eigenfunctions of the Hamiltonian (\ref{schrod_independent_gravity}):
  \[
        \chi_n(x) = \Theta(x)\frac{\mathrm{Ai}(x-E_n)}{\mathrm{Ai}'(-E_n)} .
 \]
Writing $\Psi(x,0) = \sum_n c_n \chi_n(x)$, the problem is reduced to finding an expression of the coefficients
 \[
        c_n = \braket{\chi_n}{\psi} = \frac{1}{(2\pi\zeta^2)^{\frac{1}{4}}}\int_0^{\infty}\mathrm{d}x~\chi_n^*(x) \, \mathrm{e}^{-\frac{(x-h)^2}{4\zeta^2}},
 \]
    where the asterisk denotes complex conjugation.

When the width $\zeta$ of the Gaussian is small enough with respect to $h$, the lower bound of the integral can be replaced by $-\infty$ and the $c_n$ have an analytical expression:
\begin{align} \nonumber
        c_n &= \frac{1}{(2\pi\zeta^2)^{\frac{1}{4}}\mathrm{Ai}'(-E_n)}\int_{-\infty}^{+\infty}\mathrm{d}x ~ \mathrm{Ai}(x-E_n)\mathrm{e}^{-\frac{(x-h)^2}{4\zeta^2}}\\ \nonumber
        & = \frac{2\zeta}{(2\pi\zeta^2)^{\frac{1}{4}}\mathrm{Ai}'(-E_n)} \int_{-\infty}^{+\infty}\mathrm{d}u ~ \mathrm{Ai}(2\zeta u +h -E_n)\mathrm{e}^{-u^2}\\ \nonumber
        & = \frac{(8\pi\zeta^2)^{\frac{1}{4}}}{\mathrm{Ai}'(-E_n)}\mathrm{Ai}(h - E_n +\zeta^4) \exp{\zeta^2\qty(h-E_n+\frac{2}{3}\zeta^4)} ,
    \end{align}
which is just the expression of Eq. \eqref{c_n_gravity}.
Note that we used the following identity:
\[
        \int_{-\infty}^{+\infty} \mathrm{d}u~ \mathrm{e}^{-u^2}\mathrm{Ai}(2au+b) = \sqrt{\pi}\mathrm{e}^{a^2b+\frac{2}{3}a^6}\mathrm{Ai}(b+a^4).
\]

\end{appendices}

\newpage
\bibliography{biblionelson}% common bib file
%%\input nelson-born1.bbl

\end{document}